\documentclass{appolb}

\usepackage{graphics,graphicx}
\usepackage{amsmath, amssymb}
\usepackage{multirow}
\usepackage{color}
\usepackage{verbatim}
\usepackage{hyperref}
\usepackage[normalem]{ulem}  
\usepackage{ulem}
\usepackage{color}
\usepackage{notoccite}
\usepackage{cancel}
\usepackage{mathtools}
\usepackage{epstopdf}

\renewcommand\sout{\bgroup \color{blue} \ULdepth=-.5ex \ULset}
\def\slashchar#1{\setbox0=\hbox{$#1$}  
\dimen0=\wd0     
\setbox1=\hbox{/} \dimen1=\wd1  
\ifdim\dimen0>\dimen1   
\rlap{\hbox to \dimen0{\hfil/\hfil}} 
#1     
\else     
\rlap{\hbox to \dimen1{\hfil$#1$\hfil}} 
/      
\fi}

\newcommand{\dd}{\mathrm{d}}

\begin{document}
\title{Hybrid quark-hadron equation of state for multi-messenger astronomy%
\thanks{Presented at Criticality in QCD and the Hadron Resonance Gas, 29-31 July 2020}%
}
\author{Micha\l{} Marczenko$^{\thanks{e-mail:michal.marczenko@uwr.edu.pl}}$
   \address{
   Institute of Theoretical Physics, University of Wroc\l{}aw, PL-50204 Wroc\l{}aw, Poland}
}
\maketitle
\begin{abstract}
We present the recent progress of the hybrid quark-meson-nucleon model for multi-messenger astronomy that unifies the thermodynamics of quark and hadronic degrees of freedom. The obtained equation of state is in accordance with nuclear matter properties at saturation density and with the flow constraint from heavy ion collision experiments. The mass-radius relations and tidal deformabilities for compact stars are calculated and compared with the latest astrophysical observations. The phase diagram of the isospin-symmetirc matter is studied as well in terms of the higher-order cumulants of the net-baryon number. 
\end{abstract}
\PACS{PACS numbers come here}

\section{Introduction}
\label{sec:introduction}

	The equation of state (EoS) is one of the key observables characterizing properties of matter under extreme conditions. In the context of strongly interacting matter, the EoS encodes information on the phase structure and the phase diagram of quantum chromodynamics (QCD).

	At finite temperature and low net-baryon density, the EoS, obtained in ab initio calculations within lattice QCD (LQCD), exhibits a smooth crossover from hadronic matter to a quark-gluon plasma, which is linked to the restoration of the chiral symmetry and color deconfinement~\cite{Bazavov:2018mes}. LQCD provides the state-of-the-art results describing properties of the low baryon-density QCD matter that are also successfully used for the interpretation of data obtained from the relativistic heavy ion collisions (HIC)~\cite{Bazavov:2020bjn}. At low temperature and high density the progress is driven mostly due to discoveries of various two-solar mass NSs~\cite{Demorest:2010bx, Antoniadis:2013pzd, Fonseca:2016tux, Cromartie:2019kug}.
 
	The mechanism of color confinement and its relation to the chiral symmetry breaking are of major importance in probing the hadron-quark phase transition, although it is nontrivial to embed their interplay into a single effective theory. For this reason, despite its serious shortcomings, the conventional approach is to use separate effective models for the nuclear and quark matter phases (two-phase approaches) with a priori assumed first-order phase transition, typically associated with simultaneous chiral and deconfinement transitions~\cite{Bastian:2015avq}.

	In this contribution, we briefly present our recent studies based on the hybrid quark-meson-nucleon (QMN) model~\cite{Benic:2015pia, Marczenko:2017huu, Marczenko:2018jui, Marczenko:2019trv, Marczenko:2020jma, Marczenko:2020wlc}. The model has the characteristic feature that the chiral symmetry is restored within the hadronic phase by lifting the mass splitting between chiral partner states, before the quark deconfinement takes place. Quark degrees of freedom are included on top of hadrons, but their unphysical onset is prevented at low densities. The hybrid QMN model naturally embeds the interplay between the quark confinement and the chiral symmetry breaking into a single field-theoretical framework, which makes the phenomena inherently connected, in contrast to the common two-phase approach. In this write-up, we present our results for the isospin-symmetric and isospin-asymmetric matter. 

\section{Hybrid quark-meson-nucleon model}\label{sec:hybrid_qmn}

  In this section, we briefly introduce the hybrid QMN model for the chiral symmetry restoration and deconfinement phase transitions. The hybrid QMN model is composed of the baryonic parity doublet~\cite{Detar:1988kn} and mesons as in the Walecka model, as well as quark degrees of freedom as in the standard linear sigma model. The spontaneous chiral symmetry breaking yields the mass splitting between the two baryonic parity partners, while it generates the entire mass of a constituent quark. In this work, we consider a system with $N_f=2$; hence, relevant for this study are the positive-parity nucleons, i.e., proton ($p_+$) and neutron ($n_+$), and their negative-parity partners, denoted as $p_-$ and $n_-$, as well as the up ($u$) and down ($d$) quarks. The fermionic degrees of freedom are coupled to the chiral fields $\left(\sigma, \boldsymbol\pi\right)$, the isosinglet vector-isoscalar field ($\omega_\mu$), and the vector-isovector field ($\boldsymbol \rho_\mu$). The important concept of statistical confinement is realized in the hybrid QMN model by introducing a medium-dependent modification of the particle distribution functions.

  The thermodynamic potential of the hybrid QMN model in the mean-field approximation reads
  \begin{equation}\label{eq:thermo_pot_iso}
   \Omega = \sum_{x=p_\pm,n_\pm,u,d}\Omega_x + V_\sigma + V_\omega + V_\rho + V_b \textrm.
  \end{equation}
  where the summation goes over the fermionic degrees of freedom. The potentials $V_i$ are the commonly used mean-field potentials~\cite{Marczenko:2020jma}. The spin degeneracy factor, $\gamma_x$ for nucleons is $\gamma_\pm=2$ for both positive- and negative-parity states, while the spin-color degeneracy factor for up and down quarks is $\gamma_q=2\times 3 = 6$. The kinetic part, $\Omega_x$, reads
  \begin{equation}\label{eq:thermokin}
   \Omega_x = \gamma_x \int\frac{\dd^3p}{\left(2\pi\right)^3} T \left[\ln\left(1-n_x\right) + \ln\left(1-\bar n_x\right)\right]\textrm,
  \end{equation}
  where the functions $n_x$ and $\bar n_x$ are the modified Fermi-Dirac distributions for nucleons
  \begin{subequations}\label{eq:cutoff_nuc}
  \begin{align}
         n_\pm &= \frac{1}{1+e^{\beta \left(E_\pm - \mu_\pm\right)}} \theta \left(\alpha^2 b^2 - \boldsymbol p^2\right) \textrm,\\
    \bar n_\pm &= \frac{1}{1+e^{\beta \left(E_\pm + \mu_\pm\right)}}\theta \left(\alpha^2 b^2 - \boldsymbol p^2\right) 
  \end{align}
  \end{subequations}
  and for quarks
  \begin{subequations}\label{eq:cutoff_quark}
  \begin{align}
        n_q &=\frac{1}{1+e^{\beta \left(E_q - \mu_q\right)}} \theta \left(\boldsymbol p^2-b^2\right) \textrm,\\
   \bar n_q &=\frac{1}{1+e^{\beta \left(E_q + \mu_q\right)}} \theta \left(\boldsymbol p^2-b^2\right) \textrm,
  \end{align}
  \end{subequations}
  respectively. The model embeds the concept of statistical confinement through the modified Fermi-Dirac distribution functions, where $b$ is the expectation value of an auxiliary scalar field $b$ and $\alpha$ is a dimensionless model parameter. As demonstrated in Refs.~\cite{Benic:2015pia,Marczenko:2017huu}, the parameter $\alpha$ plays also a crucial role in tuning the order of the chiral phase transition. $\beta$ is the inverse temperature, and the dispersion relation $E_x = \sqrt{\boldsymbol p^2 + m_x^2}$. The effective chemical potentials for $p_\pm$ and $n_\pm$ are defined as
  \begin{subequations}\label{eq:u_eff_had_iso}
  \begin{align}
    \mu_{p_\pm} &= \mu_B - g^N_\omega\omega - \frac{1}{2}g^N_\rho \rho + \mu_Q\textrm,\\
    \mu_{n_\pm} &= \mu_B - g^N_\omega\omega + \frac{1}{2}g^N_\rho \rho\textrm.
  \end{align}
  \end{subequations}
  The effective chemical potentials for up and down quarks are given by
  \begin{subequations}\label{eq:u_effq}
  \begin{align}
    \mu_u &= \frac{1}{3}\mu_B - g^q_\omega \omega - \frac{1}{2}g^q_\rho \rho + \frac{2}{3}\mu_Q\textrm,\\
    \mu_d &= \frac{1}{3}\mu_B - g^q_\omega \omega + \frac{1}{2}g^q_\rho \rho - \frac{1}{3}\mu_Q\textrm.
  \end{align}
  \end{subequations}
  In Eqs.~\eqref{eq:u_eff_had_iso}~and~\eqref{eq:u_effq}, $\mu_B$, $\mu_Q$ are the baryon and charge chemical potentials, respectively. 

  Because the nature of the repulsive interaction among quarks and their coupling to the $\omega$ and $\rho$ mean fields are still far from consensus, we account for the uncertainty in the theoretical predictions. To this end, we treat the couplings $g^q_\omega$ and $g^q_\rho$ as free parameters, 
  \begin{align}
    g_\omega^q &= \chi g_\omega^N\textrm, \\
    g_\rho^q &= \chi g_\rho^N \textrm,
  \end{align}
  where $\chi$ is a dimensionless parameter. 

  The effective masses of the chiral partners, $m_{p_\pm} = m_{n_\pm} \equiv m_\pm$, are given by
  \begin{equation}\label{eq:doublet_masses}
    m_\pm = \frac{1}{2} \left[ \sqrt{\left(g_1+g_2\right)^2\sigma^2+4m_0^2} \mp \left(g_1 - g_2\right)\sigma \right] \textrm.
  \end{equation}
  The positive-parity nucleons are identified as the positively charged and neutral $N(938)$ states, i.e., proton ($p_+$) and neutron ($n_+$). Their negative-parity counterparts, denoted as $p_-$ and $n_-$ are identified as $N(1535)$~\cite{Tanabashi:2018oca}. From Eq.~(\ref{eq:doublet_masses}), it is clear that the chiral symmetry breaking generates only the splitting between the two masses. When the chiral symmetry is restored, the masses become degenerate with a common finite mass $m_\pm\left(\sigma=0\right) = m_0$, which reflects the parity doubling structure of the \mbox{low-lying} baryons. Following our previous studies~\cite{Benic:2015pia,Marczenko:2017huu,Marczenko:2018jui,Marczenko:2019trv,Marczenko:2020jma}, we choose a rather large value, $m_0=700$~MeV.

  The quark effective mass, $m_u = m_d \equiv m_q$, is linked to the sigma field as 
  \begin{equation}\label{eq:mass_quark}
    m_q = g_q \sigma \textrm.
  \end{equation}
  We note that in contrast to the baryonic parity partners (cf. Eq.~\eqref{eq:doublet_masses}), quarks become massless as the chiral symmetry gets restored.

  The numerical values of the model parameters are taken from~\cite{Marczenko:2020jma}. In-medium profiles of the mean fields are obtained by extremizing the thermodynamic potential~in Eq.~\eqref{eq:thermo_pot_iso}. The allowed range for the $\alpha$ parameter is $\alpha b_0 = 300 - 450~$MeV~\cite{Benic:2015pia,Marczenko:2017huu}, where $b_0$ denotes the vacuum expectation value of the $b$-field. Following our previous works, we choose four representative values within that interval: $\alpha b_0 = 350,~370,~400,~450~$MeV.

\section{Results}\label{sec:results}

  \begin{figure}[t]
  \begin{center}
    \includegraphics[width=.7\linewidth]{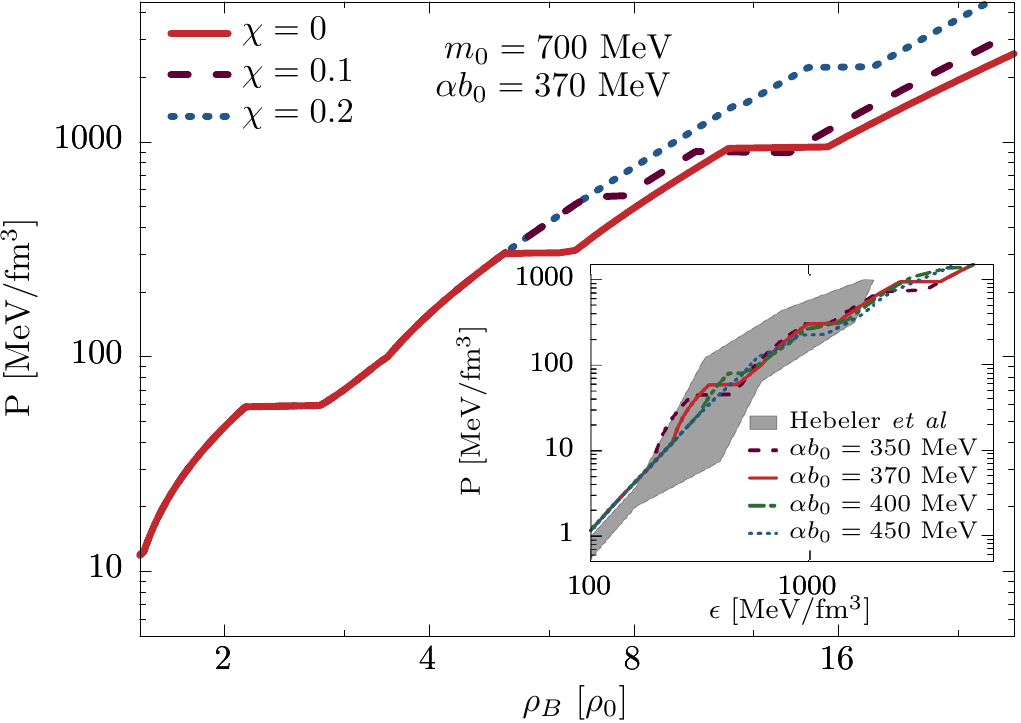}
    \caption{Thermodynamic pressure, $P$, under the NS conditions, as a function of the energy density, $\epsilon$, for $m_0 = 700~$MeV and $\chi = 0$, and four representative values of the parameter $\alpha$ (see text for details). The phase transitions are seen as plateaux of constant pressure. The gray shaded region marks the constraint obtained by~\cite{Hebeler:2013nza}.}
    \label{fig:p_e}
  \end{center}
  \end{figure}

	In Fig.~\ref{fig:p_e}, we show the EoSs under the neutron-star conditions for $m_0=700~$MeV and $\alpha b_0=370~$MeV and different values of the repulsive quark-vector coupling $\chi$. Shown EoSs feature a common first-order chiral phase transition. For comparison, the EoSs obtained for the remaining values of the parameter $\alpha$ are also shown in the vicinity of the chiral phase transitions. As the density increases, the EoSs feature another two sequential jumps in baryon density. The first is associated with the onset of the down quark, and the second is associated with the onset of the up quark, after which the EoS is composed solely of quarks. Due to the effect of the finite quark-vector coupling the onset of both quarks is systematically shifted toward higher densities when compared to the case with vanishing coupling. Consequent extension of the hadronic branch of the EoS is exhibited. In this case, the EoSs remain the same up to the point where the down quark appears. Such separation of the chirally broken and the deconfined phase might indicate the existence of a quarkyonic phase~\cite{McLerran:2008ua}. We note that the class of equations of state obtained in the hybrid QMN model is in accordance with the flow constraint from heavy ion collisions~\cite{Marczenko:2020jma}.

  \begin{figure*}[t]
  \begin{center}
    \includegraphics[width=0.7\linewidth]{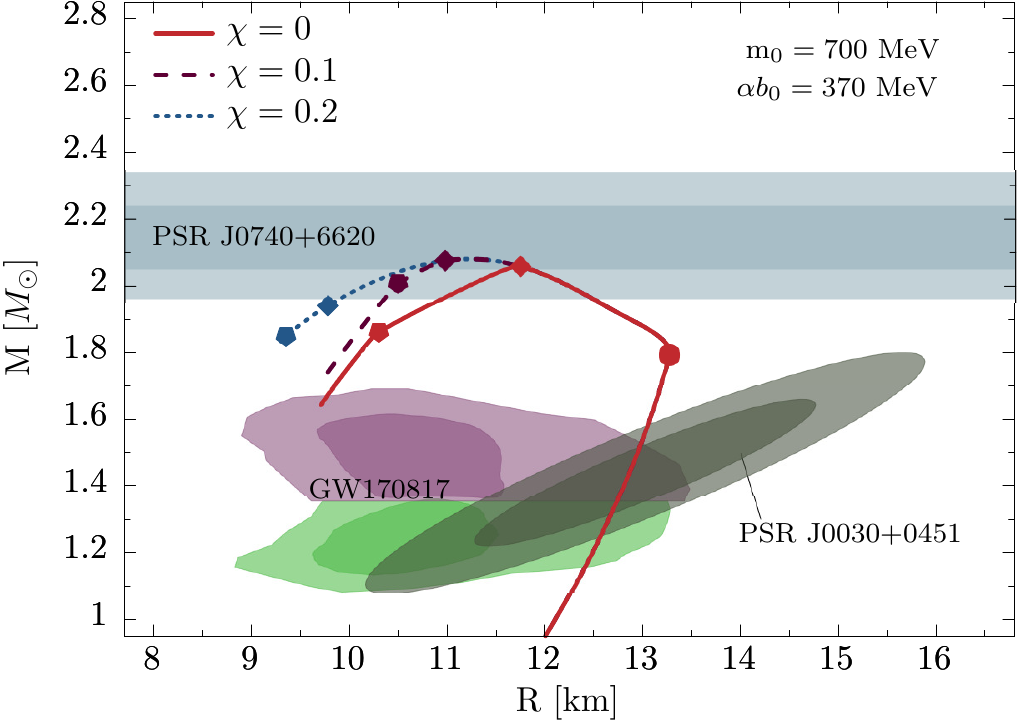}
    \caption{Mass-radius sequences for compact stars as solutions of the TOV equations for $m_0=700$~MeV, for $\alpha b_0=370~$MeV.}
    \label{fig:m_r}
  \end{center}
  \end{figure*}

   The extension of the hadronic branch of an EoS due to a finite value of the quark vector-interaction $\chi$ is also reflected in the corresponding mass-radius sequence. This is shown in Fig.~\ref{fig:m_r}. In the figure, the circles show the coexistence of the chirally broken and chirally restored phases. The onsets of up and down quarks are marked by pentagons and diamonds, respectively. The inner (outer) gray band shows the 68.3\%~(95.4\%) credibility regions for the mass of PSR J0740+6620~\cite{Cromartie:2019kug}. The inner (outer) green and purple bands show 50\%~(90\%) credibility regions obtained from the recent GW170817~\cite{Abbott:2018exr} event for the low- and high-mass posteriors. The inner (outer) black region corresponds to the mass and radius constraint at 68.2\% (95.4\%) obtained for PSR J0030+0451 by the group analyzing NICER X-ray data~\cite{Miller:2019cac}. Interestingly, the maximal mass is always reached within the hadronic branch of the sequence. For \mbox{$\chi=0$}, this happens just before the density jump associated with the onset of down quark is reached. The appearance of the down quark makes the matter too soft to sustain from the gravitational collapse. For \mbox{$\chi=0.1$} and $0.2$, the hadronic branch extends beyond the density at which the maximal mass is reached and becomes gravitationally unstable. Eventually, when down and up quark are sequentially populated, the matter is still not stiff enough to sustain from the collapse and turn into an additional family of stable hybrid compact stars. For $\chi=0.2$, the hadronic branch extends even further, however the maximal mass stays the same. Thus, we conclude that a further increase of the quark-vector coupling does not support the maximal-mass constraint.

  \begin{figure}
    \centering\includegraphics[width=0.7\linewidth]{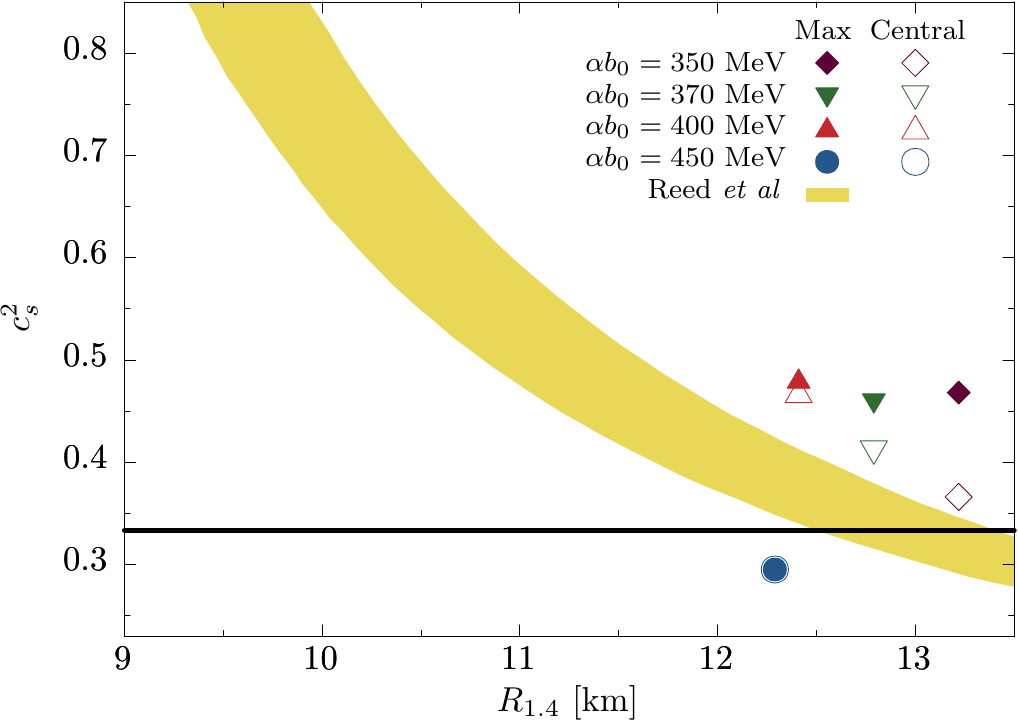}
    \caption{The maximal value of the square of the speed of sound, $c_s^2$, within a $1.4M_\odot$ NS, as a function of its radius, $R$. The results are obtained for $m_0 = 700~$MeV, and four representative values of the parameter $\alpha$.}\label{fig:m_r_cs2}
  \end{figure}
	
	In~\cite{Reed:2019ezm}, the interplay between constraints from high-mass measurements and gravitational-wave observations was used to derive a lower limit on the speed of sound in a $1.4M_\odot$ NS within a class of simplistic constant-speed-of-sound (CSS) EoSs. The constraint is shown in Fig.~\ref{fig:m_r_cs2} as a function of radius, $R_{1.4}$, of $1.4M_\odot$ NS (yellow-shaded region). The speed of sound monotonically decreases as $R_{1.4}$ increases. In the figure, we also show the conformal value $c_s^2=1/3$ (black horizontal line), the maximal values of $c_s^2$ (filled symbols) within $1.4M_\odot$ obtained for each parametrization in the hybrid QMN model, together with corresponding central values (open symbols). We note that for all obtained EoSs the $1.4~M_\odot$ NS is realized in the confined phase. For $\alpha b_0=350,~370,~400~$MeV, the values are not only above the conformal limit, but they also lie above the constraint. Notably, the maximal values of the speed of sound are obtained at densities where the stiffening of the EoS set in. On the other hand, the maximal value of the speed of sound for $\alpha b_0=450~$MeV does not exceed the conformal value. In this case, $c_s^2$ rises monotonically even beyond the central density of the $1.4M_\odot$ NS, and the stiffening sets in at higher densities. Seemingly, sufficient stiffening of the EoS at densities just above the saturation density is required in order to comply with the constraint from Ref.~\cite{Reed:2019ezm}. In the hybrid QMN model, it is provided through the dynamical mechanism of confinement which strength in linked to the density. The inclusion of the statistical confinement has important implications already at densities before the quarks are deconfined. This may have important phenomenological implications for the study of multi-messenger astronomy and heavy ion collisions (HIC)~\cite{Sasaki:2019jyh}.

    \begin{figure}[t]
    \begin{center}
      \includegraphics[width=0.7\linewidth]{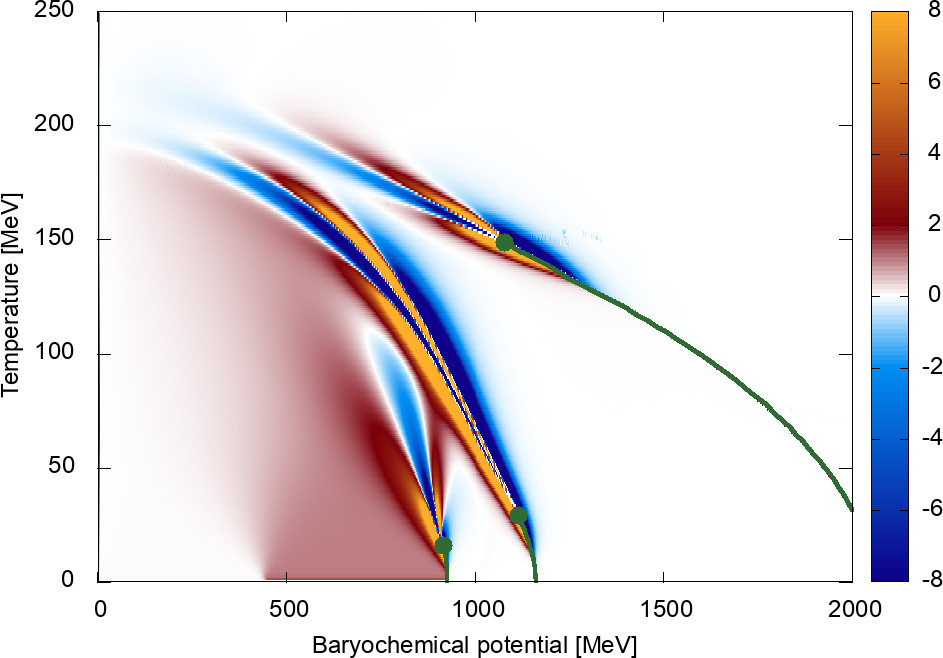}
      \caption{The isospin-symmetric phase diagram in terms of the ratio of the fifth and second-order cumulants of the net-baryon number. The green lines indicate first-order transitions. Green dots mark the critical points.}
      \label{fig:phase_diag}
    \end{center}
    \end{figure}

	In Fig~\ref{fig:phase_diag}, we show the phase diagram obtained for $m_0=700$, $\alpha b_0=370~$MeV, $\chi=0$ in terms of the ratio of the fifth- and the second-order net-baryon number cumulants, $\chi^B_5 / \chi_2^B$. The transitions are well separeated at low temperatures. At higher $T$, the transitions become less separated and the interplay between them becomes more apparent. This is particularly interesting at intermediate temperature range, where the remnant crossover signals of the liquid-gas and chiral phase transitions overlap. We leave the analysis of the highly non-trivial structure of higher-order cumulants as our future task~\cite{to_be_done}.

\section{Conclusions}\label{sec:conclusions}

	In this contribution, we have utilized the hybrid quark-meson-nucleon (QMN) model to quantify the equation of state (EoS) of cold and dense matter. The model unifies the thermodynamics of quark and hadronic degrees of freedom. The interplay between the quark confinement and the chiral symmetry breaking is embedded in a dynamical way into a single unified framework. Within this approach, we have systematically investigated the EoS of cold and dense asymmetric matter under NS conditions. We have constructed the mass-radius relations based on solutions of the Tolman-Oppenheimer-Volkoff (TOV) equations.

	We have shown that the model complies with modern constraints from multi-messenger astronomy. In particular, we analyzed a possible occurrence of exotic matter in the NS core. We have shown that the transition to pure quark matter is likely to appear in the part of the stellar sequence that is already gravitationally unstable. We note that the inclusion of color superconducting quark matter phases may allow the existence of hybrid quark-hadron stars due to the lowering of the onset mass for deconfinement while fulfilling the maximum mass constraint. We have also argued that a rapid increase of pressure is required at densities inside a $1.4M_\odot$ NS. In the hybrid QMN model, such stiffening is naturally connected to the dynamical mechanism of confinement which strength in linked to the density. This result highlights the fact that the confinement plays a crucial role in the phenomenology of matter under extreme conditions, even at densities smaller than the density at which the system undergoes a hadron-to-quark phase transition.

\medskip

	The author acknowledges fruitfull collaboration and discussions with D.~Blaschke, K.~Redlich, and C.~Sa\-sa\-ki. This work was partly supported by the Polish National Science Center (NCN), under Preludium Grant No. UMO-2017/27/N/ST2/\-01\-973.

\medskip


\end{document}